\documentclass[aps,prb,twocolumn,showpacs,groupedaddress]{revtex4-1}
\usepackage{graphicx}% Include figure files
\usepackage{dcolumn}% Align table columns on decimal point
\usepackage{color}
\usepackage{bm}% bold math
\usepackage{ulem}

\newcommand{\Ns}{\ensuremath{N_\mathrm{s}}}

\newcommand{\Tc}{\ensuremath{T_\mathrm{c}}}
\newcommand{\DI}{\ensuremath{D_\mathrm{I}}}

\newcommand{\redsout}[1]{}

\begin{document}
\title{Loop algorithm for classical Heisenberg models with spin-ice type degeneracy} 
\author{Hiroshi Shinaoka}
\altaffiliation[Present address: ]{Nanosystem Research Institute, AIST, Tsukuba 305-8568, Japan}
\affiliation{Institute for Solid State Physics, University of Tokyo, Kashiwanoha, Kashiwa, Chiba, 277-8581, Japan}
\author{Yukitoshi Motome} 
\affiliation{Department of Applied Physics, University of Tokyo, 7-3-1 Hongo, Bunkyo-ku, Tokyo 113-8656, Japan}
\date{\today}

%%%-----------------------------------------------------------------
\begin{abstract}
	In many frustrated Ising models, a single-spin flip dynamics is frozen out at low temperatures compared to the dominant interaction energy scale because of the discrete ``multiple valley" structure of degenerate ground-state manifold. This makes it difficult to study low-temperature physics of these frustrated systems by using Monte Carlo simulation with the standard single-spin flip algorithm. A typical example is the so-called spin ice model, frustrated ferromagnets on the pyrochlore lattice. The difficulty can be avoided by a global-flip algorithm, the loop algorithm, that enables to sample over the entire discrete manifold and to investigate low-temperature properties. We extend the loop algorithm to Heisenberg spin systems with strong easy-axis anisotropy in which the ground-state manifold is continuous but still retains the spin-ice type degeneracy. We examine different ways of loop flips and compare their efficiency. The extended loop algorithm is applied to the following two models, a Heisenberg antiferromagnet with easy-axis anisotropy along the $z$ axis, and a Heisenberg spin ice model with the local $\langle 111 \rangle$ easy-axis anisotropy. For both models, we demonstrate high efficiency of our loop algorithm by revealing the low-temperature properties which were hard to access by the standard single-spin flip algorithm. For the former model, we examine the possibility of order-from-disorder and critically check its absence. For the latter model, we elucidate a gas-liquid-solid transition, namely, crossover or phase transition among paramagnet, spin-ice liquid, and ferromagnetically-ordered ice-rule state.
\end{abstract}


\maketitle
%%%-----------------------------------------------------------------
%%%  Introduction
%%%-----------------------------------------------------------------
\section{Introduction}
Geometrically frustrated systems have attracted much attention because of fascinating phenomena arising from competing interactions~\cite{Diep05}. Frustration prevents simultaneous optimization of all interaction energies, which suppresses long-range ordering and may lead to a macroscopic number of energetically (nearly-)degenerate ground states. Such ground-state manifold plays a decisive role in low-temperature($T$) physics under the influence of quantum/thermal fluctuations and external perturbations. It is highly important to clarify the structure of the manifold and to take the statistical average over the entire manifold for understanding low-$T$ properties in frustrated systems. 

\begin{figure}[!]
 \centering
 \includegraphics[width=.35\textwidth,clip]{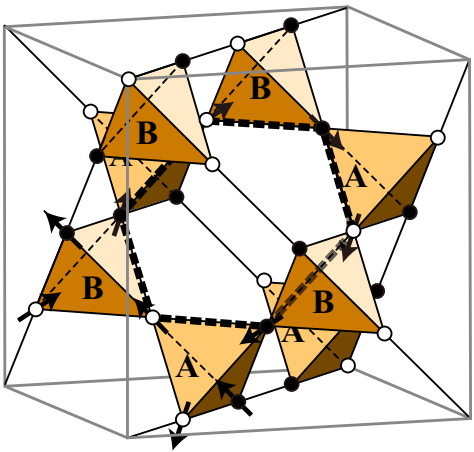}
 \caption{(color online). The pyrochlore lattice composed of a three-dimensional network of corner-sharing tetrahedra. A 16-site cubic unit cell is shown. A and B represent upward and downward tetrahedra, respectively. Ising spins are denoted by arrows. Each spin axis is along the local $\langle 111 \rangle$ axis, which goes from the site to the center of a neighboring tetrahedron. Black (filled) circles represent spins pointing inward, while white (open) circles represent spins pointing outward in terms of type-A tetrahedra. The hexagon with a bold dashed line denotes one of the shortest loops on which a flip of all spins (colors) transforms an ice-rule state to another ice-rule state. See the text for details.}
 \label{fig:model}
\end{figure}

As a typical example, we consider antiferromagnets with classical spins on the pyrochlore lattice. The pyrochlore lattice is a three-dimensional frustrated structure given by a corner-sharing network of tetrahedra, as shown in Fig.~\ref{fig:model}. When the system has the Heisenberg $O(3)$ symmetry and the exchange interaction is limited to nearest-neighbor sites, any long-range ordering does not occur and the ground-state manifold has continuous macroscopic degeneracy~\cite{Reimers92}. The manifold is identified by a collection of local constraints, that is, the summation of spin vectors on four vertices should vanish in every tetrahedron. This condition is underconstraint and leaves two angles undetermined in each tetrahedron, resulting in the continuous macroscopic degeneracy~\cite{Moessner98a,Moessner98b}. A similar zero-sum local constraint can be found in an Ising ferromagnet on the pyrochlore lattice with local cubic $\langle 111 \rangle$ axes, the so-called spin ice model~\cite{Harris97,Ramirez99}, which is equivalent to a pyrochlore Ising antiferromagnet with a global anisotropy axis~\cite{Anderson56}. In this case, the local constraint enforces two spins pointing inward and two spins pointing outward in every tetrahedron, as exemplified in Fig.~\ref{fig:model}. This two-in two-out constraint is called the ice rule because of an analogy to the constraint on positions of protons in hexagonal ice~\cite{Bernal33,Pauling35}. The ice rule is also underconstraint, leading to the disordered ground-state with macroscopic degeneracy. In this case, the manifold has a discrete nature because of the Ising spin degree of freedom. Intermediate type manifolds, namely not discrete but not fully $O(3)$, also appear in variants of pyrochlore antiferromagnets, such as anisotropic Heisenberg models~\cite{Bramwell94} and bilinear-biquadratic Heisenberg models~\cite{Shannon10}.

In the nearest-neighbor antiferromagnetic Heisenberg model, all energetically-degenerate spin configurations in the continuous manifold are connected by continuous changes of spin directions without energy cost because of the continuous structure of the manifold. This is indeed observed in classical Monte Carlo (MC) studies of the pyrochlore Heisenberg antiferromagnets; a standard single-spin flip update is efficient to sample over the entire manifold down to very low $T$ compared to the exchange energy scale $J$. In contrast, for discrete Ising-type or continuous but strongly anisotropic manifolds, degenerate spin configurations are separated by large energy barriers of the order of $J$, and a single-spin flip does not work at low $T \ll J$. In fact, in the spin ice model with long-range dipole interactions, it is hard to clarify low-$T$ properties by single-spin flip MC calculations, and there was a controversy about the possibility of long-range ordering~\cite{Hertog00, Bramwell01,Melko01}.

To overcome such difficulty coming from ``multiple valley'' structure of the degenerate manifold, it is necessary to consider a global flip which connects different degenerate configurations. For the spin ice models, this is achieved by introducing the loop-flip algorithm, in which one reverses all Ising spins on a specific closed loop passing through tetrahedra~\cite{Rahman72, Yanagawa79, Barkema98,Melko01}; the loop is chosen so that the Ising spins are inward and outward alternatively along the loop (in/outward is defined on one of two different types of tetrahedra, say, type-A tetrahedra in Fig.~\ref{fig:model}). This loop flip enables to transform an ice-rule state to another ice-rule state bypassing the energy barriers. Melko \textit{et al.} applied the loop algorithm and successfully observed symmetry breaking emergent from the ice-rule manifold in the dipolar spin-ice model without severe dynamical freezing~\cite{Barkema98,Melko01}. The loop algorithm has been successfully applied to study the low-$T$ properties of spin-ice type Ising models~\cite{Isakov04,Jacob05, Jaubert10}.

The loop algorithm is well defined for systems with discrete Ising-type spins. However, when the discreteness is relaxed and spins can fluctuate around the anisotropy axis, it becomes nontrivial how to define the loop with alternating in/outward spins. Moreover, it is also unclear how thermal spin fluctuations affect the acceptance rate of the loop flip update.
As long as the degenerate manifold retains a ``multiple valley'' structure with large energy barriers, a single-spin flip update becomes inefficient and some global flip is indispensable for taking the statistical average in an ergodic way. Such problems are encountered in many systems, e.g., frustrated Heisenberg models with strong easy-axis anisotropy or with large biquadratic interactions. To elucidate low-$T$ properties in this type of frustrated systems, it is desired to establish a global flip update applicable to systems with continuous but ice-rule type degenerate manifold. Note that the situation has an aspect similar to the Wolff's extension~\cite{Wolff89} of the Swendsen-Wang cluster algorithm~\cite{Swendsen87} for conventional unfrustrated magnets.

In this paper, we extend the loop algorithm to Heisenberg spin systems with strong easy-axis anisotropy in which the ground-state manifold is continuous but retains the ice-rule type ``multiple valley'' structure. We describe how to define the loop in such systems, and discuss two different types of loop flip. Because of the continuous degrees of freedom of Heisenberg spins, the way of flipping is not unique and its efficiency depends on the method. To demonstrate the efficiency of the extended loop algorithm, we apply the method to two different models, the Heisenberg antiferromagnetic model with easy-axis anisotropy and the Heisenberg spin ice model. For the former, we discuss the possibility of order-from-disorder by thermal fluctuations. For the latter, we map out the phase diagram which includes a spin-ice like liquid state as well as a ferromagnetically-ordered ice-rule state. 

This paper is organized as follows. In Sec. II, after reviewing the loop algorithm for Ising spin systems, we extend it to Heisenberg models with easy-axis anisotropy. In Sec. III and IV, we apply the extended loop algorithm to the pyrochlore antiferromagnet with the $z$-axis anisotropy and the spin-ice type ferromagnet with $\langle 111 \rangle$ anisotropy, respectively. Summary is given in Sec. V.

%%%-----------------------------------------------------------------
%%%  Algorithm
%%%-----------------------------------------------------------------
\section{Monte Carlo Algorithm}

\subsection{Loop algorithm for Ising spin systems}

Before considering an extension of the loop algorithm to Heisenberg spin systems, here we briefly review the loop algorithm for Ising spin models with ice-rule type degeneracy~\cite{Barkema98, Melko01}. To generalize the following discussion, we assign black and white to two degrees of freedom of Ising spins in an appropriate manner. For example, for the spin ice model~\cite{Harris97,Ramirez99}, black and white represent inward and outward spins in terms of type-A tetrahedra, respectively, as shown in Fig.~\ref{fig:model}. For the antiferromagnetic Ising model~\cite{Anderson56}, black and white simply correspond to up and down spins, respectively. Then the loop flip consists of two steps; first, we identify a closed loop which consists of alternating alignment of black and white sites, and then we try to flip all Ising spins on the loop. 

\begin{figure}[t]
 \centering
 \resizebox{0.35\textwidth}{!}{\includegraphics{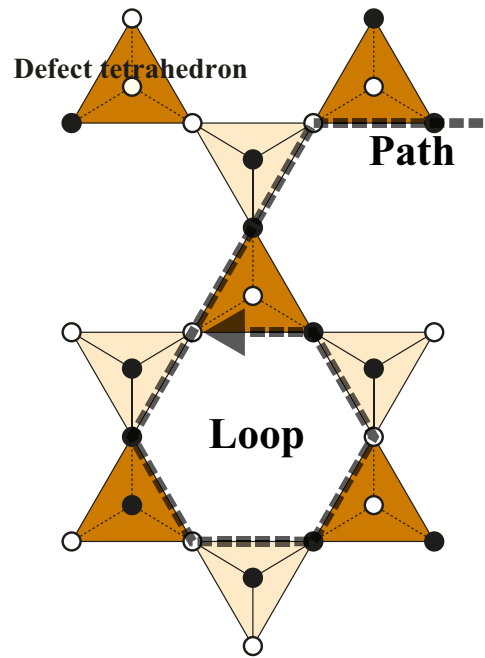}}
 \caption{(color online). Schematic picture for a loop construction by tracing a path through ice-rule tetrahedra. The path is made of alternating black and white sites. For simplicity, the figure shows a $\langle 111 \rangle$ kagome layer with connected tetrahedra. The path is denoted by a dashed line, and its lower part represents an example of a closed loop. }
 \label{fig:loop}
\end{figure}

When all tetrahedra satisfy the ice rule as in the ice-rule ground states, it is trivial to construct a loop of alternating black and white sites. At finite $T$, however, thermal fluctuations induce ``defect tetrahedra" in which the ``two-in two-out" ice-rule condition is violated and ``three-in(out) one-out(in)" or ``four-in(out)" configuration is realized (see Fig.~\ref{fig:loop}). To maintain detailed balance in the loop flip, a loop must be contructed so that it does not involve defect tetrahedra on it.

Two methods were proposed for the loop construction; one is called the long loop algorithm~\cite{Rahman72, Barkema98} and the other is the short loop algorithm~\cite{Rahman72, Yanagawa79, Melko01}. In the present study, we focus on the short loop algorithm. The procedure is the following (Fig.~\ref{fig:loop}):

\begin{enumerate}
 \item First, we randomly choose a tetrahedron which satisfies the ice-rule local constraint, namely, a tetrahedron with two black and two white sites.
 \item We move to one of its four neighboring tetrahedra which also satisfies the ice rule, and mark the site shared by the two tetrahedra as the first site (the origin of path). If all neighboring tetrahedra are defect, we abort the attempt to construct a loop and go back to the step 1.
 \item Then we move to a tetrahedron which satisfies the ice rule out of three neighboring tetrahedra (except for the tetrahedra visited just before), and mark the shared site as the second site. We repeat the procedure and extend the path of marked sites which consists of an alternation of black and white sites. As in the step 2, once all three tetrahedra are defect, we abort the attempt and go back to the step 1.
\item If one of three neighboring tetrahedra has already been visited, we move to it and make a closed loop by deleting the dangling tail of the path. 
\end{enumerate}
The procedure is slightly modified in the steps 3 and 4 from the original short loop algorithm~\cite{Rahman72, Yanagawa79, Melko01}. In the original version, a move to neighboring tetrahedra is always completely random (except for the one visited just before); it does not avoid defect tetrahedra (once a defect tetrahedron is chosen, the attempt is aborted), and does not choose a previously-visited tetrahedron selectively. The above modifications enhance the efficiency by reducing the possibility to fail a loop construction, with satisfying the detailed balance in MC calculations. 

After the construction of a closed loop, all the colors on the loop are reversed simultaneously. This corresponds to a flip of all Ising spins on the loop: $\vec{S}_i \rightarrow - \vec{S}_i$. When the system has nearest-neighbor interactions only, the loop flip does not change the total energy; in other words, the flip is always accepted (rejection free update) in the MC sampling. When there are residual interactions such as farther-neighbor interactions, the loop flip is accepted with the probability depending on the total energy change by the standard Metropolis algorithm.

In the long loop algorithm~\cite{Rahman72, Barkema98}, a closed loop is formed only when the path returns to the initial site. This can generate a longer loop and its flip causes a bigger change of configurations. However, in general, it takes more CPU time to construct a longer loop. Moreover, a flip of a longer loop leads to a larger energy change and a lower acceptance rate. (For nearest-neighbor models, there is no energy cost and the long loop algorithm can be efficient.) Therefore, the short loop algorithm is more efficient than the long loop algorithm in general.

At finite $T$, it is clear that the loop flip update does not satisfy ergodicity because it can change neither the spin configurations in defect tetrahedra nor the number of defect tetrahedra. It is, therefore, necessary to use the loop flip together with another update for retaining the ergodicity. This is easily achieved by introducing the standard single-spin flip in MC samplings.

\subsection{Extension of the loop algorithm to anisotropic Heisenberg spin systems}
In this section, we extend the loop algorithm to Heisenberg models with easy-axis anisotropy. We start with a Hamiltonian of a general form: 
\begin{equation}
\mathcal{H}= -J \sum_{i,j} %\sum_{\langle i,j \rangle} 
\vec{S}_i \cdot \vec{S}_j - %D_\mathrm{I} 
\DI \sum_i %\sum_{i=1}^{\Ns} 
\left(\vec{S}_i \cdot \vec{\alpha}_i\right)^2, \label{eq:ham-h}
\end{equation}
where $\vec{S}_i$ denotes a classical Heisenberg spin at site $i$ on the pyrochlore lattice (we take $|\vec{S}_i|=1$) and $\DI$ $(>0)$ is the single-ion easy-axis anisotropy. The unit vector $\vec{\alpha}_i$ defines the easy axis on site $i$. In the case of antiferromagnets with easy-axis anisotropy along the $z$ axis, which are extensions of the Ising antiferromagnet~\cite{Anderson56}, we set $\vec{\alpha}_i = (0,0,1)$ for all the sites. While, in the case of ferromagnets with the local $\langle 111 \rangle$ anisotropy, which are natural extensions of the spin ice model~\cite{Harris97,Ramirez99}, we set $\vec{\alpha}_i$ to the direction connecting the centers of neighboring tetrahedra from type B to A.

To define a loop for the model given by Eq.~(\ref{eq:ham-h}), 
we assign black and white colors to sites at which $\vec{S}_i \cdot \vec{\alpha}_i \ge 0$ and $\vec{S}_i \cdot \vec{\alpha}_i < 0$, respectively. Based on this definition, we can construct a closed loop with alternating black and white sites by following the short loop algorithm in Sec. II A.
Then, we try to reverse all colors on the loop. However, this loop flip procedure is not unique in the Heisenberg spin case. To choose an efficient method, careful consideration on the energy change is necessary as discussed below. 

\begin{figure}[!]
 \centering
 \resizebox{0.35\textwidth}{!}{\includegraphics{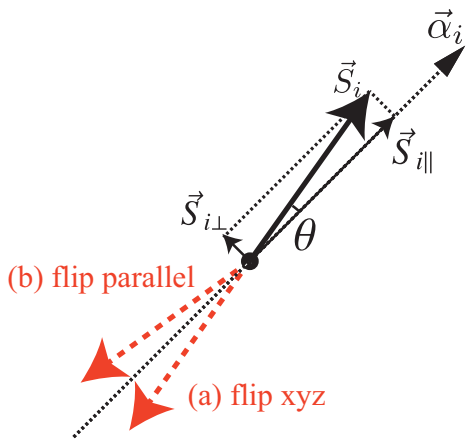}}
 \caption{(color online). Two different ways to flip black and white for the spin $\vec{S}_i$ at site $i$: (a) \textit{flip xyz} and (b) \textit{flip parallel}. See the text for details.}
 \label{fig:flip}
\end{figure}

A natural extension of the loop flip in the Ising case is to reverse all three Cartesian components of $\vec{S}_i$ on the loop as $\vec{S}_i \rightarrow - \vec{S}_i$, which we call \textit{flip xyz}, as illustrated in Fig.~\ref{fig:flip}. This \textit{flip xyz} changes the energy by
\begin{eqnarray}
 \Delta E &=& 2J \sum_{i \in \mathrm{loop}} 
 \sum_{j \notin \mathrm{loop}} \vec{S}_j \cdot \vec{S}_i \nonumber\\
 &=& 2J \sum_{i \in \mathrm{loop}}
 \sum_{j \notin \mathrm{loop}} \vec{S}_j \cdot (\vec{S}_{i\parallel} + \vec{S}_{i\perp})\nonumber \\
 &=& \Delta E_\parallel + \Delta E_\perp \label{eq:flipxyz}, 
\end{eqnarray}
where 
\begin{eqnarray}
 \Delta E_\parallel &=&  2J \sum_{i \in \mathrm{loop}}
 \sum_{j \notin \mathrm{loop}} \vec{S}_j \cdot \vec{S}_{i\parallel}, \label{eq:Eparallel}\\
 \Delta E_\perp &=&  2J \sum_{i \in \mathrm{loop}}
 \sum_{j \notin \mathrm{loop}} \vec{S}_j \cdot \vec{S}_{i\perp}, \label{eq:Eperp}
\end{eqnarray}
and $\vec{S}_{i\parallel}$ ($\vec{S}_{i\perp}$) is the component of $\vec{S}_i$ parallel (perpendicular) to $\vec{\alpha}_i$ (see Fig.~\ref{fig:flip}). For models with nearest-neighbor interactions which retain ice-rule degeneracy in the ground state, one might expect that this update is always accepted in the limit of $T \rightarrow 0$; this is naively expected since thermal fluctuations vanish and all the ice-rule configurations with spins parallel to the easy axes become energetically degenerate. However, this is not the case. At low $T$, spins fluctuate around the easy axes by angles of the order of $T$, and hence, the energy change by the flip is estimated as
\begin{eqnarray}
 \Delta E_\parallel &\propto& \sum_{i \in \mathrm{loop}} \theta_i^2 \propto T^2, \label{eq:Eparallel-T2}\\
 \Delta E_\perp &\propto&  \sum_{i \in \mathrm{loop}} \theta_i \propto T, \label{eq:Eperp-T}
\end{eqnarray}
where $\theta_i$ denotes a deviation angle of spin $i$ from the easy axis. Because the acceptance rate is given by $\mathrm{min} \{1, \exp(-\Delta E/T)\}$, thermal fluctuations are irrelevant for the flip of $\vec{S}_{i\parallel}$ in the sense that $\lim_{T\rightarrow 0} \exp(- \Delta E_\parallel/T) = 1$. On the other hand, thermal fluctuations are relevant for the flip of $\vec{S}_{i\perp}$ since $\lim_{T\rightarrow 0} \exp(- \Delta E_\perp/T)  < 1$. Therefore, \textit{flip xyz} does not become rejection free in the limit of $T\rightarrow 0$ even for nearest-neighbor models with ice-rule degeneracy. 

This consideration suggests a better way of a global update of spin configurations on the loop. That is a flip of only parallel components $S_{i\parallel}$ as $\vec{S}_i \rightarrow \vec{S}_i  - 2 (\vec{S}_i \cdot \vec{\alpha}_i) \vec{\alpha}_i$. This \textit{flip parallel} changes the energy only by $\Delta E_\parallel  \propto T^2$ at low $T$, and hence, is expected to become rejection free as $T \to 0$ for nearest-neighbor models with ice-rule degeneracy. The efficiency of these two updates, \textit{flip xyz} and \textit{flip parallel}, will be compared in numerical simulations in the following sections.

As mentioned above, to retain the ergodicity, we use the loop flip together with the single-spin flip. One MC step consists of single-spin flips, followed by loop flips with either \textit{flip xyz} or \textit{flip parallel}.
In the single-spin flips, we randomly choose a new spin state on the unit sphere for each spin following a procedure proposed by Marsaglia~\cite{Marsaglia72}.
The loop flips are repeated until the number of tetrahedra visited exceeds the number of lattice sites. In our implementation, the procedure of loop flips takes CPU time comparable to that of a sweep of the lattice sites by single-spin flips.

In the single-spin flips, we note that the completely random choice of a new spin direction leads to a low acceptance rate at low $T$, and a high acceptance rate is retained when restricting the new spin state within a small angle $\delta$ around the original spin direction. However, the high acceptance rate does not directly mean the high efficiency for the spin-ice type models, because such single-spin flips cannot retain the ergodicity at low $T$: small local fluctuations around the easy axes hardly lead to a global update between different spin-ice states at low $T \ll J$. On the other hand, the random single-spin flips give a physically-important measure: its steep suppression signals formation of the spin-ice type manifold. This will be demonstrated for the two models in Sec. III and IV.

%%%-----------------------------------------------------------------
%%%  Bencghmark1
%%%-----------------------------------------------------------------
\section{Application to pyrochlore Heisenberg antiferromagnets with easy-axis anisotropy}
In this section, we apply the extended loop algorithm to the Heisenberg antiferromagnetic model with easy-axis anisotropy on the pyrochlore lattice. After introducing the model in Sec. III A, we demonstrate the efficiency of loop flips in MC simulations in Sec. III B. In Sec. III C, we discuss the possibility of order-from-disorder phenomenon in comparison with related models.

\subsection{Model}
We consider an antiferromagnetic Heisenberg model with easy-axis anisotropy along the $z$ axis on the pyrochlore lattice. The model is given by taking $\vec{\alpha}_i = (0,0,1)$ for all sites in the model (\ref{eq:ham-h}). For simplicity, we take account of nearest-neighbor interactions only. The Hamiltonian is given by
\begin{equation}
\mathcal{H}= -J \sum_{\langle i,j \rangle} \vec{S}_i \cdot \vec{S}_j - \DI \sum_i {\left({S}_i^z\right)}^2. \label{eq:ham-h2}
\end{equation}
Here we consider the antiferromagnetic case $J<0$, and set an energy scale as $|J|=1$, i.e., $J=-1$.
In the following MC calculations, we denote the linear dimension of the system measured in the cubic unit cell by $L$. Namely, the total number of spins in the system $\Ns$ is given by $16L^3$. Hereafter, we employ periodic boundary conditions.

In the limit of $\DI\rightarrow \infty$, this model reduces to the antiferromagnetic Ising model on the pyrochlore lattice. The ground state of the Ising model has the spin-ice type degeneracy, i.e., all spin configurations with two-up and two-down spins on every tetrahedron are energetically degenerate~\cite{Anderson56}. It is known that the system does not show any phase transition at finite $T$ in this case~\cite{Liebmann86}. For $0 < \DI < \infty$, the situation at $T=0$ does not change; the ground state has the same macroscopic degeneracy. An interesting question is whether the degeneracy is lifted by thermal fluctuations in the Heisenberg spin model defined by Eq.~(\ref{eq:ham-h2}). In other words, the question is whether an order-from-disorder phenomenon takes place in this model. To answer this question, it is necessary to investigate the low-$T$ properties, much lower than $|J|$ where the ice-rule manifold is gradually formed. 

\subsection{Demonstration of loop flip}
\begin{figure}[!]
 \centering
 \resizebox{0.4\textwidth}{!}{\includegraphics{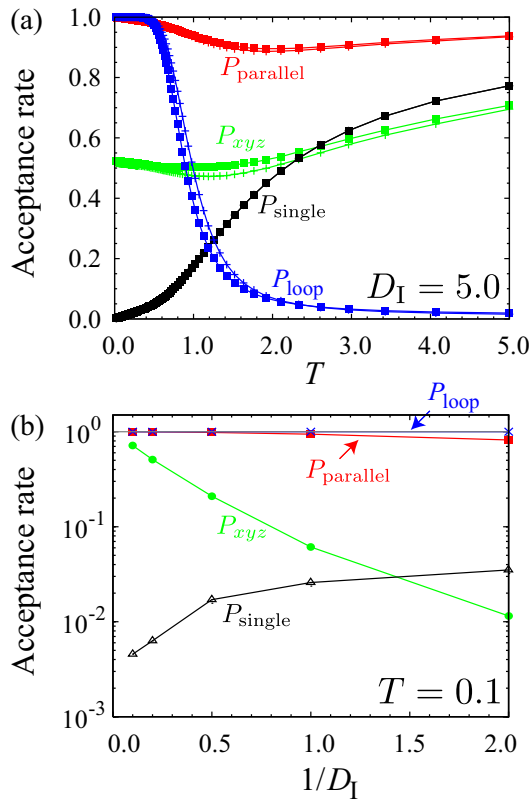}}
 \caption{(color online). (a) Temperature dependence of the acceptance rate of the single-spin flip ($P_{\text{single}}$), the probability of formation of closed loops ($P_{\text{loop}}$), the acceptance rates of flip of a formed loop by \textit{flip xyz} ($P_{xyz}$) and by \textit{flip parallel} ($P_{\text{parallel}}$). The data are calculated for the model (\ref{eq:ham-h2}) at $\DI=5.0$. The data for the system sizes $L=2$ and $L=4$ are denoted by crosses and filled squares, respectively. (b) $\DI$ dependence of $P_{\text{parallel}}$, $P_{xyz}$, $P_{\text{loop}}$, and $P_{\text{single}}$ at $T=0.1$ for $L=2$.}
 \label{fig:z}
\end{figure}

We investigate finite-$T$ properties of the model (\ref{eq:ham-h2}) by MC simulation with the loop algorithm extended in Sec. II B. 
At low $T$ compared to $|J|$, spin configurations are gradually enforced to satisfy the ice rule, and the acceptance rate of the single-spin flip, $P_{\text{single}}$, is suppressed. This is demonstrated in Fig.~\ref{fig:z}(a) for $\DI=5.0$: $P_{\text{single}}$ rapidly decreases at $T \sim |J|$ and vanishes almost linearly in $T$ in the limit of $T \to 0$. On the contrary, the acceptance rate of loop flips increases at low $T$. As shown in Fig.~\ref{fig:z}(a), the probability that a closed loop is successfully formed, $P_\mathrm{loop}$, steeply increases below $T \sim |J|$, indicating that almost all tetrahedra start to follow the ice rule below this temperature. At the same time, the acceptance rate of flips of a formed loop gradually increases at low $T < |J|$ and remains finite; here, $P_{xyz}$ and $P_{\text{parallel}}$ are the rate for \textit{flip xyz} and \textit{flip parallel}, respectively. The total acceptance rate of the loop flip is given by the product as $P_\mathrm{loop} \times P_{xyz}$ or $P_\mathrm{loop} \times P_{\text{parallel}}$, and it sharply increases at $T < |J|$, compensating the decrease of $P_{\text{single}}$. 

As clearly indicated in Fig.~\ref{fig:z}(a), the acceptance rate of \textit{flip parallel} is always larger than that of \textit{flip xyz}. In particular, $P_{\text{parallel}}$ approaches 1 (rejection free) as $T \to 0$, whereas $P_{xyz}$ goes to a smaller value $\sim 0.5$. The reduction of $P_{xyz}$ becomes larger for smaller anisotropy $\DI$. This is demonstrated at $T=0.1$ in Fig.~\ref{fig:z}(b); $P_{xyz}$ decreases almost exponentially with $1/\DI$ and decreases to about 0.01\ at $\DI=0.5$. 
On the other hand, $P_{\text{parallel}}$ is almost independent of $\DI$ and remains rejection free as $T \to 0$ in the wide range of  $\DI$.  
Note that the single-spin flip does not work efficiently even for weak anisotropy, e.g., for $\DI=0.5$, $P_{\text{single}} \simeq 0.03$ at $T=0.1$. Therefore, \textit{flip parallel} retains higher efficiency than \textit{flip xyz} and compensates the low efficiency of the single-spin flip over a wide range of $\DI$.
These behaviors are consistent with the argument in Sec. II B, and demonstrate the advantage of \textit{flip parallel} at low $T$.

\begin{figure}[!]
 \centering
 \resizebox{0.4\textwidth}{!}{\includegraphics{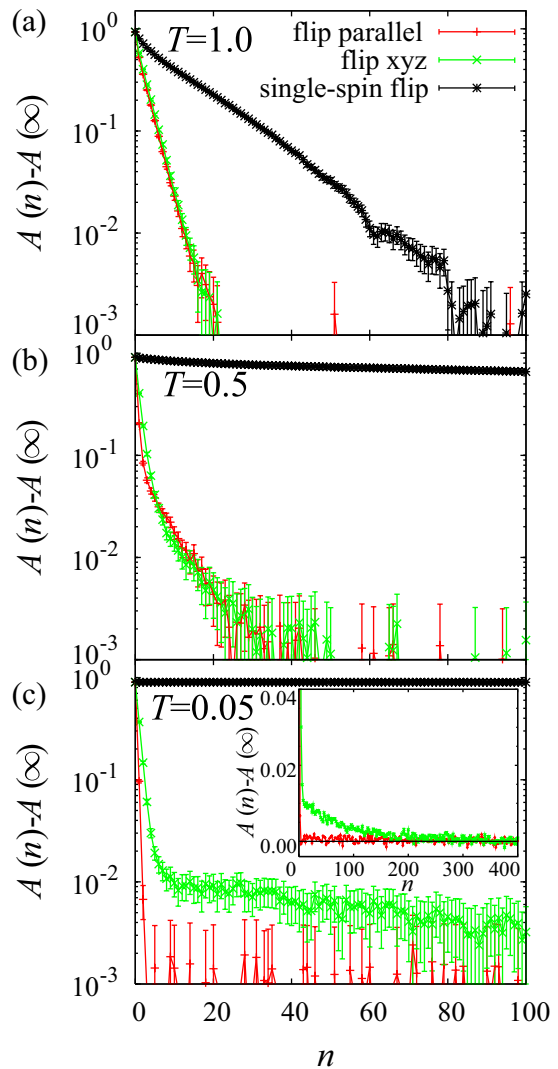}}
 \caption{(color online). Autocorrelation functions for spin configurations [Eq.~(\ref{eq:autocorr})] calculated at (a) $T=1.0$, (b) $T=0.5$, and (c) $T=0.05$. The data are at $\DI = 5.0$ for $L=2$. The offset $A(\infty)$ was obtained by averaging $A(n)$ in the range of $ 5000 \le n \le 10000$. The inset of (c) shows the same data as in (c) in a wider range of $n$.}
 \label{fig:auto-correlation}
\end{figure}

To further demonstrate efficiency of the loop flips, we calculate the autocorrelation function of spin configurations. 
Here the autocorrelation function is defined for an interval of $n$ MC steps in the form
\begin{eqnarray}
 A(n) &=& \frac{1}{\Ns} \left|\sum_i \vec{S}_i (n_0) \cdot \vec{S}_i (n_0 + n)\right|,
 \label{eq:autocorr}
\end{eqnarray}
where $\Ns$ is the number of lattice sites.
The results at $T=1.0, 0.5, 0.05$ are shown in Fig.~\ref{fig:auto-correlation}. We average the data over independent $10^4$ runs after $n_0 = 1\times 10^4$ thermalization. The decay of the autocorrelation function for the single-spin flip dynamics rapidly becomes slower at low $T$: The autocorrelation remains almost 1 for $n<100$ at the lowest $T=0.05$. This clearly indicates the freezing of single-spin flip at low $T$. In contrast, the loop-flip dynamics exhibits no signature of freezing down to the lowest $T$ in the present calculations. In particular, even at $T=0.05$, the autocorrelation vanishes rapidly for \textit{flip parallel}; it becomes smaller than $0.01$ for $n\ge 3$, indicating the autocorrelation time is ${\cal O}(1)$. On the other hand, as clearly indicated in the inset of Fig.~\ref{fig:auto-correlation}(c), the loop-flip dynamics with \textit{flip xyz} exhibits a slower residual relaxation for $n>10$. This slow relaxation may be due to enhanced spin fluctuations around thermally-induced ``defect tetrahedra'', which lower the efficiency of \textit{flip xyz}. The \textit{flip parallel} does not severely suffer from such fluctuations and keeps the high efficiency down to the lowest $T$.

\subsection{Absence of order from disorder}
As mentioned in Sec. III A, in the Ising limit $\DI \to \infty$, the model does not show any phase transition, and the ground state has continuous macroscopic degeneracy~\cite{Anderson56,Liebmann86}.
%and the ground state has the ice-rule type discrete degeneracy~\cite{Anderson56,Liebmann86}.
On the other hand, in the Heisenberg limit, i.e., $\DI = 0$, the model shows no transition down to $T=0$ and the ground state has a continuous degeneracy~\cite{Reimers92,Moessner98a,Moessner98b}. However, it was pointed out that, for the planar $XY$ type anisotropy with $\DI < 0$, although the system suffers from a continuous degeneracy in the ground state, thermal fluctuations select a subset from the continuous manifold and induces a first-order transition to a conventional N\'{e}el order~\cite{Bramwell94}. Furthermore, on the 2D kagome lattice (a $\langle 111 \rangle$ plane in the pyrochlore lattice), $XXZ$ models with Ising-type exchange anisotropy also exhibit a thermally-driven phase transition by selecting a subset from an ice-rule type discrete manifold of the ground state~\cite{Kuroda95}. These results are examples of the so-called order-from-disorder phenomena appearing in the systems with anisotropy. 

Motivated by these previous studies, we here examine the possibility of order-from-disorder in the present model with Ising type anisotropy $0 < \DI < \infty$. We calculate the specific heat $C$ and the uniform magnetic susceptibility $\chi_0$ by MC simulation with the single-spin flip and the loop flip of \textit{flip parallel}. 
The specific heat $C$ and the uniform magnetic susceptibility $\chi_0$ are calculated by
\begin{eqnarray}
 C  &=& \frac{1}{\Ns}\frac{\langle \mathcal{H}^2 \rangle - \langle \mathcal{H} \rangle^2}{T}
\end{eqnarray}
and 
\begin{eqnarray}
 \chi_0 &=& \frac{1}{3\Ns}\frac{\langle M^2 \rangle}{T},
\end{eqnarray}
respectively, where $\langle \cdots \rangle$ denotes a thermal average. Here the square of total magnetization $M^2$ is given by $M^2 = \sum_{\mu=x,y,z} \left(\sum_{i} S_{i}^{\mu}\right)^2$. Numbers of MC steps for thermalization, $N_\mathrm{th}$, and for sampling, $N_\mathrm{samp}$, are $(N_\mathrm{th}, N_\mathrm{samp}) = (1\times 10^4, 1\times 10^5)$ for $L=2$ and $L=4$, $(5\times 10^4, 5\times 10^5)$ for $L=5$, respectively. The data are averaged over four independent MC runs to estimate statistical errors by variance of average values in the runs. 

\begin{figure}[!]
 \centering
 \resizebox{0.4\textwidth}{!}{\includegraphics{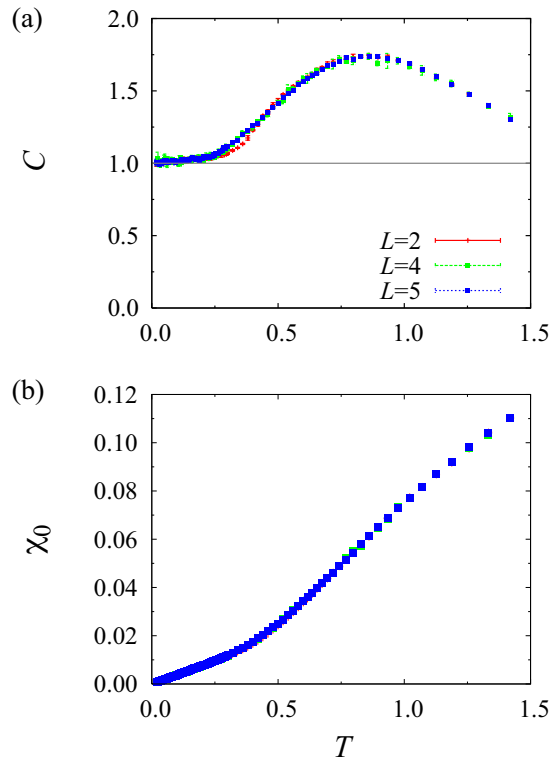}}
 \caption{(color online). Temperature dependences of (a) the specific heat $C$ and (b) the uniform magnetic susceptibility $\chi_0$ for the model (\ref{eq:ham-h2}) at $\DI=5.0$.} 
 \label{fig:phys-D5.0}
\end{figure}

Figure~\ref{fig:phys-D5.0} shows the results at $\DI=5.0$. Although we find a broad peak in the specific heat $C$ at $T \simeq 0.85$ corresponding to cooperative formation of ice-rule tetrahedra [see also $P_{\text{loop}}$ in Fig.~\ref{fig:z}], there is no signature of a phase transition in the specific heat and the magnetic susceptibility down to the calculated lowest temperature $T=0.02$.
The data do not show any singularity and collapse onto each other among different sizes for $L\le 5$.
This indicates the absence of order-from-disorder in the present model with Ising anisotropy $\DI > 0$. We confirmed that the situation is unchanged for several other values of positive $\DI$. The absence of order-from-disorder is seemingly consistent with a Maxwellian counting argument as follows~\cite{Moessner98b}. The pyrochlore Heisenberg model is more underconstrained than the kagome one which exhibits order-from-disorder. In fact, the Heisenberg model on the pyrochlore lattice is a marginal model in terms of order-from-disorder; namely, a simple Maxwellian counting alone cannot conclude whether the order-from-disorder occurs or not. Our numerical results show that it does not occur in the present model with easy-axis anisotropy, in contrast to the case of easy-plane anisotropy. This contrasting behavior depending on the sign of $\DI$ appears to support that the pyrochlore case is marginal.

%%%-----------------------------------------------------------------
%%%  Bencghmark2
%%%-----------------------------------------------------------------
\section{Application to Heisenberg spin ice model}
In this section, we apply the extended loop algorithm to the Heisenberg spin ice model. After introducing the model in Sec. IV A, we demonstrate the efficiency of loop flips in MC simulation in Sec. IV B. In Sec. IV C, we clarify the phase diagram, and discuss the nature of crossover and phase transition among paramagnet, spin-ice liquid, and ferromagnetically-ordered spin-ice state.

\subsection{Model}
We consider a Heisenberg ferromagnet with the local $\langle 111 \rangle$ easy-axis anisotropy (Heisenberg spin ice model), whose Hamiltonian is given by
\begin{equation}
\mathcal{H}= -J \sum_{\langle i,j \rangle} \vec{S}_i \cdot \vec{S}_j - \DI \sum_i \left(\vec{S}_i \cdot \vec{\alpha}_i\right)^2.\label{eq:ham-h3}
\end{equation}
Here $\vec{S}_i$ denotes a classical Heisenberg spin at site $i$ and $\DI~(>0)$ is the single-ion easy-axis anisotropy. The unit vector $\vec{\alpha}_i$ defines the local $\langle 111 \rangle$ easy axis on site $i$, which is along the direction connecting the center of two tetrahedra sharing the site $i$ from type B to A (see Fig.~\ref{fig:model}). For simplicity, we take account of nearest-neighbor interactions only. We consider the ferromagnetic exchange $J>0$, and take the energy unit as $J=1$.
As in the previous section, the system size is $\Ns=16L^3$ spins in the following calculations.
 
\begin{figure}[!]
 \centering
 \resizebox{0.325\textwidth}{!}{\includegraphics{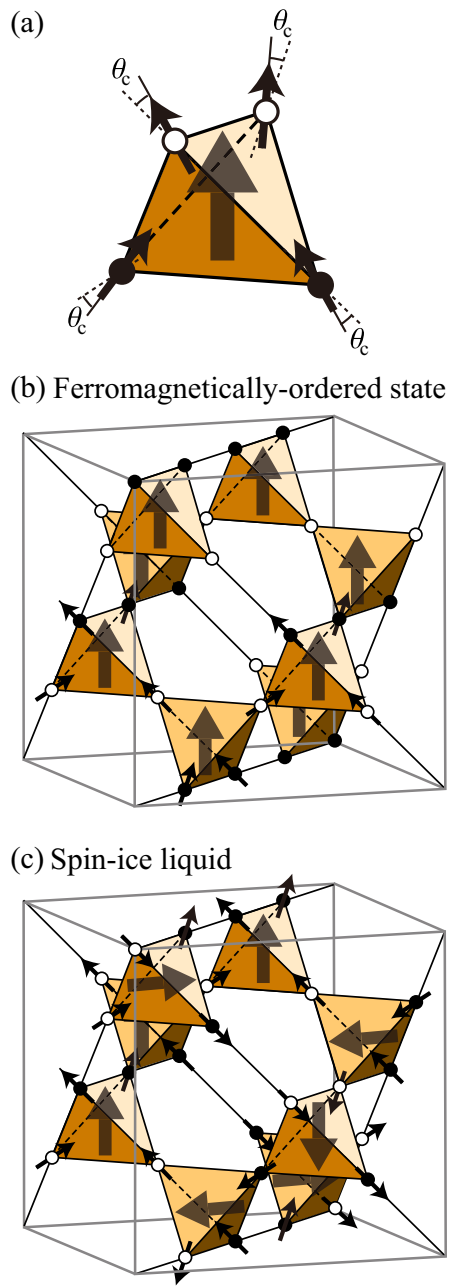}}
 \caption{(color online). (a) Schematic picture for a ground state of the Heisenberg spin ice model given by Eq.~(\ref{eq:ham-h3}) on an isolated tetrahedron. $\theta_{\text{c}}$ is a canting angle. The big arrow in the center of tetrahedron denotes a net moment by forming the canted two-in two-out spin configuration. (b) A ferromagnetically-ordered ice-rule state with aligning net magnetizations along the $z$ axis. (c) Schematic picture for a spin-ice liquid, in which each tetrahedron is in the two-in two-out state but the induced net moments are disordered among tetrahedra.}
 \label{fig:magnetic-order}
\end{figure}

The Ising counterpart of this model ($\DI=\infty$) is the nearest-neighbor spin-ice model, whose ground state retains the ice-rule degeneracy. The model does not exhibit any phase transition down to $T=0$. However, it shows a crossover at $T^* \sim {\cal O}(J)$ related with cooperative formation of the ice-rule degenerate manifold, which is signaled by a broad peak in the specific heat~\cite{Harris98,Ramirez99}.
When $0< \DI < \infty$, the system tends to gain the exchange energy by canting spins from the easy axes, as schematically shown in Fig.~\ref{fig:magnetic-order}(a). 
Through this canting, the net moments in each tetrahedron associated with the two-in two-out spin configuration are aligned among tetrahedra to minimize the exchange energy; therefore, we expect a canted ferromagnetic ground state for $0 < \DI < \infty$, as shown in Fig.~\ref{fig:magnetic-order}(b).

In fact, a recent MC study showed that the model with a positive $\DI$ exhibits a finite-$T$ phase transition~\cite{Champion02}: 
The system exhibits a transition from the high-$T$ paramagnet to 
the low-$T$ ferromagnetically-ordered ice-rule state, and 
the transition temperature $T_{\text{c}}$ increases as $\DI$ decreases. 
However, the obtained phase diagram is limited to a region with relatively-weak anisotropy, $\DI<25$.
This is because the MC study was done by using the single-spin flip algorithm 
which suffers from a severe freezing in strongly anisotropic cases. 
In the calculated region of $\DI$, 
the transition temperature $T_{\text{c}}$ is always larger than $T^*$, and 
the ordering takes place before forming the ice-rule manifold at $T^*$. 

%Hence, there remains an interesting question related with the formation of ice-rule manifold and an emergence of ordering from the manifold. 
Hence, there remains an interesting question: What happens when the ice-rule manifold is formed before the ordering? In other words, how does the formation of ice-rule manifold affect the phase transition?
In the vicinity of the Ising limit $\DI \gg J$, we expect that $T_{\text{c}}$ becomes very small and even lower than $T^*$. 
Hence, the system provides a chance to investigate the relation between a crossover from high-$T$ paramagnet (\textit{gas}) to a cooperative ice-rule state (\textit{liquid}) [Fig.~\ref{fig:magnetic-order}(c)] and the phase transition from the spin-ice liquid to the ferromagnetic ordering of the ice-ruled tetrahedra (\textit{solid}).
This is a gas-liquid-solid transition in terms of spins. 
To elucidate their nature, it is necessary to properly sample the ice-rule manifold without dynamical freezing, which will be a good target of the extended loop algorithm.

\subsection{Demonstration of loop flip}
\begin{figure}[!]
 \centering
 \resizebox{0.4\textwidth}{!}{\includegraphics{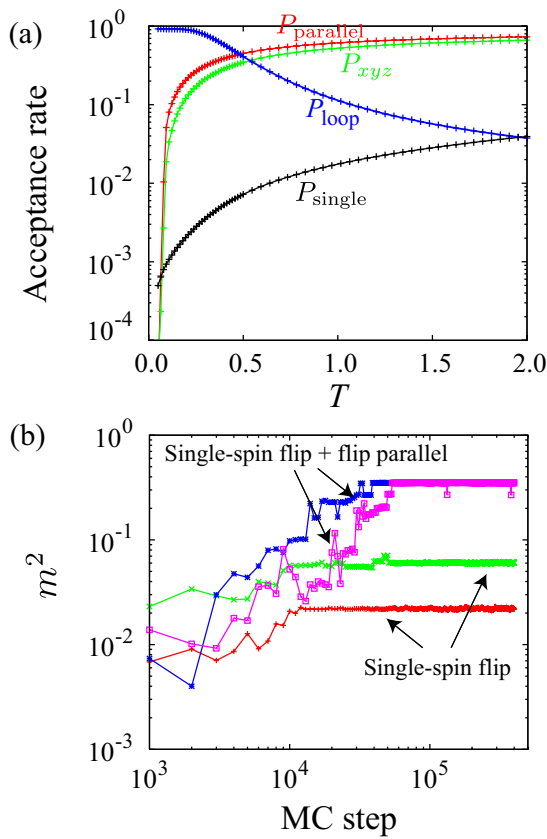}}
 \caption{(color online). (a) Temperature dependence of the acceptance rates of different updates. The notations are the same as in Fig.~\ref{fig:z}. (b) MC dynamics of the squared uniform magnetization $m^2$ in the MC runs with and without \textit{flip parallel} at $T=0.05$. The data are for $L=2$.
 Two data for each case show the results starting from different initial spin configurations.
}
 \label{fig:M2-dynamics}
\end{figure}

We demonstrate the efficiency of the loop algorithm for the strong easy-axis anisotropy, i.e., $\DI \gg J$. Figure~\ref{fig:M2-dynamics}(a) shows the comparison of the acceptance rates at $\DI=50.0$ as an example. Because of the large $\DI$, the acceptance rate of the single-spin flip, $P_{\text{single}}$, is strongly suppressed and becomes less than 1\% already at $T \sim J$ (cf. Fig.~\ref{fig:z}); 
with further decreasing $T$, $P_{\text{single}}$ is steeply reduced because of gradual formation of the ice-rule manifold.
On the contrary, the probability of loop formation, $P_\mathrm{loop}$, rapidly increases below $T \sim J$. At the same time, the acceptance rates of \textit{flip parallel} and \textit{flip xyz}, $P_\mathrm{parallel}$ and $P_{xyz}$, remain much larger compared to $P_{\text{single}}$. Hence, the loop flips are effective down to low $T$ and compensate the freezing of the single-spin-flip dynamics. It is also noted that $P_\mathrm{parallel}$ is always larger than $P_{xyz}$, being consistent with the argument in Sec. II B. In contrast to the model in Sec. III, however, $P_\mathrm{parallel}$ and $P_{xyz}$ do not approach a finite value as $T \to 0$ but exhibit a sharp drop at $T\simeq 0.16$. This is due to the ferromagnetic transition mentioned above. We will analyze the nature of the transition in more details in the next subsection.

To further demonstrate the efficiency of the loop flips at low $T$, we calculate MC dynamics of the squared magnetization $m^2$ for $L=2$ at $T=0.05$ (below the transition temperature $\Tc$). The squared magnetization $m^2$ is defined by
\begin{eqnarray}
	m^2 &\equiv& \left\langle \left| \frac{1}{N_\mathrm{s}}\sum_i \vec{S}_i \right|^2 \right\rangle.
\end{eqnarray}
As shown in Fig.~\ref{fig:M2-dynamics}(b), when we employ the single-spin flip only, MC dynamics shows a severe freezing; $m^2$ does not converge to an expected value $\sim 0.37$ (the value will be discussed later) even after $4\times 10^5$ MC steps, and moreover, the data are frozen after $\sim 10^4$ MC steps at some different values depending on initial spin configurations. On the other hand, MC dynamics is greatly accelerated by the loop flip and $m^2$ converges to the expected thermal-equilibrium value after $5\times 10^4$ MC steps. These clearly show the advantage of the loop algorithm in investigating the low-$T$ properties of the Heisenberg spin ice model for $\DI \gg J$.

\subsection{Spin-ice liquid-to-solid transition}
We investigate thermodynamic properties of the Heisenberg spin ice model using the loop flip of \textit{flip parallel} and the single-spin flip. To accelerate the MC dynamics further, particularly at very low $T$ where $P_\mathrm{parallel}$ is suppressed, we adopt the overrelaxation update~\cite{Michael87} and the exchange MC method~\cite{Hukushima96}. In the overrelaxation update, the energy change arising from the second term in Eq.~(\ref{eq:ham-h3}) is treated by the standard Metropolis algorithm. Numbers of MC steps for thermalization, $N_\mathrm{th}$, and for sampling, $N_\mathrm{samp}$, are $(N_\mathrm{th}, N_\mathrm{samp}) = (5\times 10^5, 5\times 10^5), (5\times 10^5, 5\times 10^5), (2\times 10^6, 2\times 10^6)$ for $L=2, 3, 4$, respectively. The data are averaged over four independent MC runs to estimate statistical errors by variance of average values in the runs. 

\begin{figure}[!]
 \centering
 \resizebox{0.4\textwidth}{!}{\includegraphics{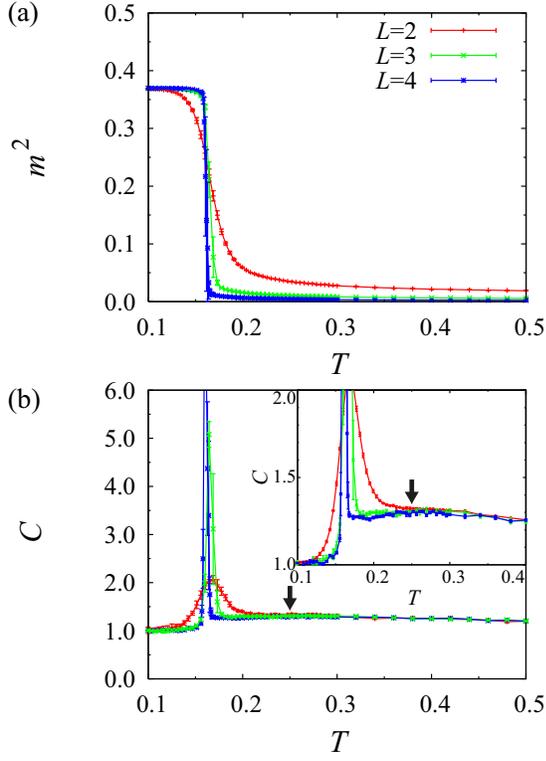}}
 \caption{(color online). Temperature dependences of (a) the squared uniform magnetization $m^2$ and (b) the specific heat $C$ at $D_\mathrm{I} = 50.0$. A broad peak in the specific heat is denoted by an arrow. The inset shows an enlarged view of the main panel.}
 \label{fig:DI50.0}
\end{figure}

First, we show $T$ dependences of the squared magnetization $m^2$ and the specific heat $C$ at $\DI=50.0$ in Fig.~\ref{fig:DI50.0}.
As shown in Fig.~\ref{fig:DI50.0}(a), $m^2$ exhibits a steep rise at $T \simeq 0.16$. This rise becomes steeper with increasing system size and almost discontinuous in the largest size $L=4$. The saturation value is $m^2 \simeq 0.37$ which is expected for the canted ferromagnetic ordered state discussed in Sec. IV A (the value will be discussed below). At the same temperature, $C$ exhibits a sharp peak, which also becomes almost discontinuous at $L=4$. These indicate that the system exhibits a first-order transition to the ferromagnetic state with a canted ice-rule spin configuration [Fig.~\ref{fig:magnetic-order}(b)]. 
These sharp singularities allow us to determine the transition temperature $\Tc$ with enough accuracy for the present purpose.

Another anomaly is found as a broad peak of $C$ at $T^* \simeq 0.25$ as shown in the inset of Fig.~\ref{fig:DI50.0}(b). Since the broad peak does not show any significant size dependence and since a similar peak is also seen in the Ising counterpart, this is a crossover from paramagnet to a cooperative ice-rule state; below  $T^*$, all tetrahedra tend to satisfy the ice-rule spin configurations but the system still remains paramagnetic for $T>\Tc$. In fact, $T^*$ coincides with the sharp rise of $P_{\text{loop}}$ in Fig.~\ref{fig:M2-dynamics}(a). We call this correlated state for $\Tc<T<T^*$ \textit{spin-ice liquid}, whose schematic picture is shown in Fig.~\ref{fig:magnetic-order}(c).

\if0
The first-order nature of the transition from this spin-ice liquid to the ferromagnetically ordered state (\textit{spin-ice solid}) is naturally understood by considering the effective degree of freedom in the ice-rule manifold. Below $T^*$, every tetrahedron almost obeys the ice rule, resulting in a net moment as shown in Fig,~\ref{fig:magnetic-order}. There are six different configurations for the two-in two-out ice rule in a single tetrahedron, which leads to an effective six-state Potts like degree of freedom. The transition at $T_{\text{c}}$ breaks this six-fold degeneracy. It can be of first order since a six-state Potts model exhibits a first-order transition in three dimensions~\cite{Wu82}.
\fi

\begin{figure}[!]
 \centering
 \resizebox{0.4\textwidth}{!}{\includegraphics{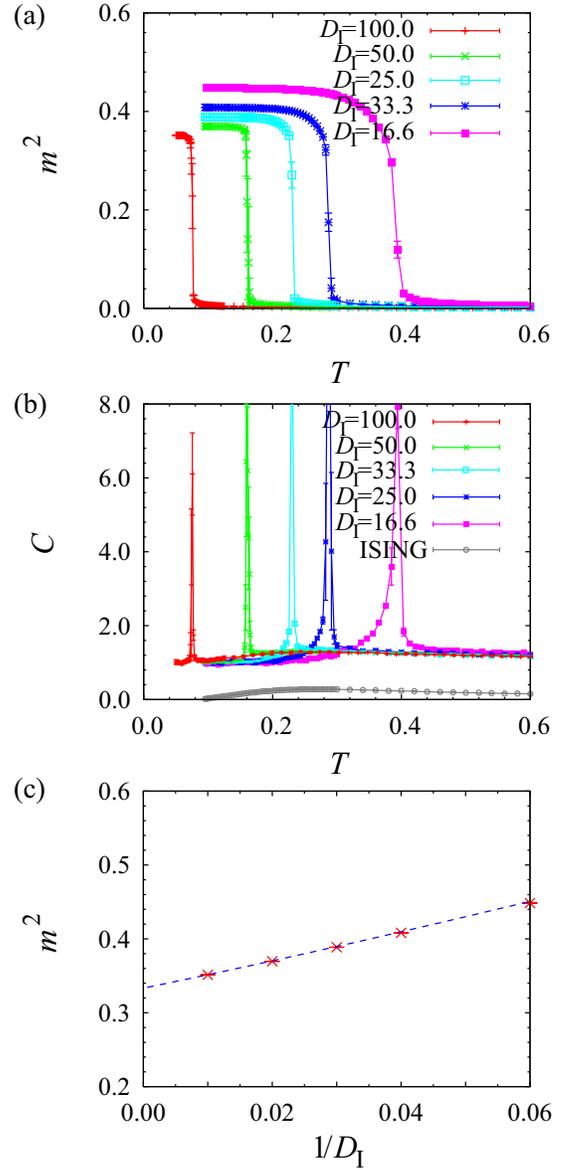}}
 \caption{(color online). Temperature dependences of (a) the squared uniform magnetization $m^2$ and (b) the specific heat $C$.
 (c) $\DI$ dependence of $m^2$ at the lowest temperature in (a);
the dotted line denotes the ground-state magnetization calculated from Eq.~(\ref{eq:m2}). The system size is $L=4$.}
 \label{fig:DI-dep}
\end{figure}
\begin{figure}[!]
 \centering
 \resizebox{0.4\textwidth}{!}{\includegraphics{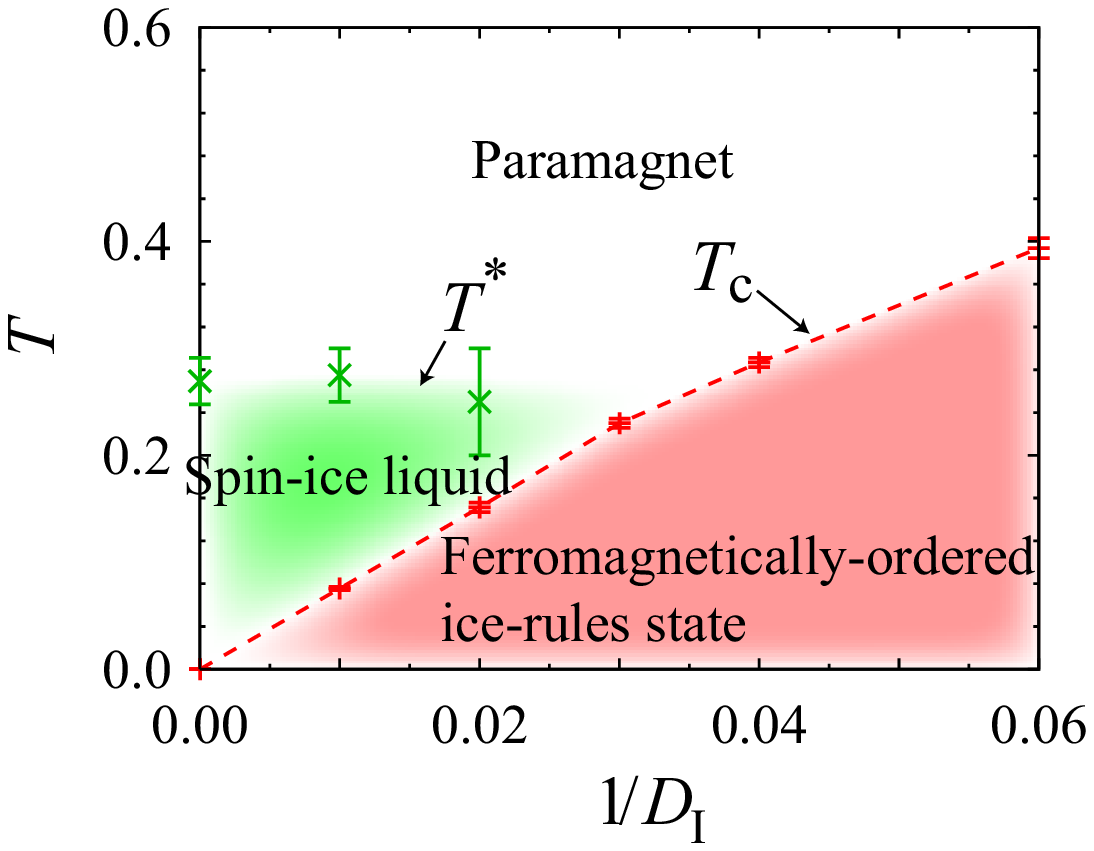}}
 \caption{(color online). $\DI$-$T$ phase diagram of the Heisenberg spin ice model given by Eq.~(\ref{eq:ham-h3}). The transition temperature $\Tc$ and the crossover temperature $T^*$ are estimated from a sharp peak and a broad one in the specific heat for $L=4$, respectively.}
 \label{fig:phase-diagram}
\end{figure}

Now we examine $\DI$ dependence of the phase transition and the crossover. Figure~\ref{fig:DI-dep} shows the calculated results of $m^2$ and $C$ for $L=4$ at $\DI = 100.0, 50.0, 33.3, 25.0, 16.6$. Numbers of MC steps for thermalization, $N_\mathrm{th}$, and for sampling, $N_\mathrm{samp}$, are $(N_\mathrm{th}, N_\mathrm{samp}) = (1\times 10^6, 1\times 10^6)$ for $\DI = 100.0, 25.0$, $(4\times 10^6, 4\times 10^6)$ for $\DI=33.3$, and $(6\times 10^4, 6\times 10^4)$ for $\DI = 16.6$. As shown in Fig.~\ref{fig:DI-dep}, $\Tc$, signaled by a sharp rise of $m^2$ and a singular peak in $C$, increases as $\DI$ decreases. On the other hand, $T^*$, at which $C$ shows a broad peak, does not strongly depend on $\DI$. As a consequence, for $1/\DI>0.04$, $T_{\text{c}}$ becomes higher than $T^*$; the spin-ice liquid state is restricted to $T_{\text{c}} < T < T^*$ for $1/\DI<0.04$. 
Note that the previous MC study was restricted to the region of $1/\DI>0.04$ where $T_{\text{c}} > T^*$ and the single-spin flip does not show severe freezing by the formation of the ice-rule manifold at $T^*$.

At the lowest $T$, $m^2$ approaches a constant whose value is dependent on $\DI$. 
By considering the ground-state energy following the discussion in Ref.~\onlinecite{Champion02}, 
the saturation value can be calculated as
\begin{eqnarray}
 m^2(T=0) &=& \frac{1}{3} \left(\sqrt{2} \sin\theta_\mathrm{c} + \cos\theta_\mathrm{c}\right)^2 \label{eq:m2},
\end{eqnarray}
where $\theta_\mathrm{c}$ is the optimal canting angle 
\begin{eqnarray}
 \theta_\mathrm{c} &=& \frac{\pi}{4} - \frac{1}{2} \arctan\frac{\sqrt{2}}{16}\left(\frac{3\DI}{J} - 4\right).
\end{eqnarray}
Note that $m^2=0$ in the Ising limit ($1/\DI=0$), while $m^2 \to 1/3$ in the limit of $1/\DI \rightarrow 0$: The anisotropy $\DI$ is a singular perturbation to the macroscopically-degenerate ground-state in the Ising case.
As shown in Fig.~\ref{fig:DI-dep}(c), $m^2$ at the lowest $T$ in our MC simulation scales well with Eq.~(\ref{eq:m2}): This demonstrates the efficiency of our algorithm down to the lowest $T$ even in the large $\DI$ region where $T_{\text{c}} < T^*$.

We summarize the results in the phase diagram in Fig.~\ref{fig:phase-diagram}. As mentioned above, $T_{\text{c}}$ grows as $1/\DI$ increases; in particular, it almost linearly increases in the small $1/\DI$ region. On the other hand, $T^*$ remains almost constant irrespective of $\DI$. Consequently, in the region of $1/\DI < 0.04$, there is a successive crossover and phase transition, i.e., a crossover from the high-$T$ paramagnet (gas) to the intermediate-$T$ spin-ice liquid at $T^*$, and a transition from the spin-ice liquid to the low-$T$ ferromagnetically-ordered spin-ice solid at $T_{\text{c}}$. In contrast, for $1/\DI > 0.04$, there is a single transition from the high-$T$ paramagnet to the low-$T$ spin-ice solid, being consistent with the previous result~\cite{Champion02}. The transition at $T_{\text{c}}$ is of first order, while the discontinuity appears to become weaker as $1/\DI$ increases~\cite{Champion02} (see Fig.~\ref{fig:DI-dep}). The phase diagram illuminates the rich physics in the Heisenberg spin ice model, including the interesting spin-ice liquid-to-solid phase transition, which has not been revealed in the previous study.

A similar liquid-to-solid transition was recently discussed for the bilinear-biquadratic Heisenberg model on the pyrochlore lattice in applied magnetic field~\cite{Shannon10}. In this model, a strong biquadratic interaction under the influence of magnetic field enforces the ice-rule like local constraint on spin configurations. Farther-neighbor interactions lift the ice-rule like degeneracy and induce a phase transition to a long-range ordered state. The phase diagram is quite similar, and the transition is of first order also in this case. 

\section{Summary}
In this paper, we have extended the loop algorithm to Heisenberg spin models with easy-axis anisotropy which have spin-ice type degeneracy in the ground state. In particular, we have examined two different ways of loop flips, \textit{flip xyz} and \textit{flip parallel}, and compared their efficiency.
By considering effects of thermal spin fluctuations around the easy axes, we have clarified that \textit{flip parallel} becomes rejection free as $T\rightarrow 0$ but \textit{flip xyz} not for models in which the ground state has macroscopic ice-rule degeneracy.

We have demonstrated the efficiency of the loop flips by performing MC simulations for two typical models, the Heisenberg antiferromagnet with easy-axis anisotropy along the $z$ axis and the Heisenberg spin ice model. By using the extended loop algorithm, we have investigated low-$T$ properties of the two models, which are hard to access by the standard single-spin flip alone. For the former model,  we have critically checked the absence of order-from-disorder phenomenon. For the latter model, we have successfully obtained the rich phase diagram involving the gas-liquid-solid like transition among the paramagnet, spin-ice liquid, and ferromagnetically-ordered ice-rule state.

Before closing this paper, we make a brief remark on the application of the extended loop algorithm to the bilinear-biquadratic Heisenberg model on the pyrochlore lattice. 
It was recently pointed out that ferromagnetic biquadratic interactions lead to formation of ice-rule type manifold and farther-neighbor bilinear interactions perturb this manifold to select an ordered state\cite{Shannon10}. In particular, under applied magnetic field, the system shows a gas-liquid-solid like transition, similar to the Heisenberg spin ice model discussed in this paper. 
The low-$T$ properties have not yet been fully clarified, because of the freezing of single-spin flips.
We expect that our extended loop algorithm works efficiently also in this model. Such extension will be reported elsewhere.

%%%-----------------------------------------------------------------
\begin{acknowledgments}
We thank T. Misawa, Y. Tomita, and T. Kato for fruitful discussions. H.S. thanks Institute for Solid State Physics for financial support. Numerical calculation was partly carried out at the Supercomputer Center, Institute for Solid State Physics, Univ. of Tokyo. This work was supported by Grant-in-Aid for Scientific Research (No. 19052008, 21340090, 22540372), Global COE Program ``Physical Sciences Frontier'', the Next Generation Super Computing Project, and Nanoscience Program, from MEXT, Japan.
\end{acknowledgments}
%%%-----------------------------------------------------------------
\bibliography{reference}
\end{document}